\documentclass[10pt]{bmc_article}

\usepackage{cite} 
\usepackage{varioref} 
\usepackage{url} 
\usepackage{ifthen} 
\usepackage{multicol}  
\usepackage[utf8]{inputenc} 
\usepackage{amsmath}
\usepackage{graphicx}
\newboolean{publ}

\newenvironment{bmcformat}{\baselineskip20pt\sloppy\setboolean{publ}{false}}{\baselineskip20pt\sloppy}

\begin{document}
\begin{bmcformat}


\title{Using Topological Data Analysis for diagnosis pulmonary embolism}
 

\author{Matteo Rucco\correspondingauthor$^1$%
     \email{Matteo Rucco\correspondingauthor - matteo.rucco@unicam.it}
    , 
     Lorenzo Falsetti$^2$%
     \email{Lorenzo Falsetti - lorenzo.falsetti@ospedaliriuniti.marche.it}
, 
     Damir Herman$^3$%
     \email{Damir Herman - damir@ayasdi.com}
	, 
     Tanya Petrossian$^3$%
     \email{Tanya Petrossian - tanya@ayasdi.com}
	,
	Emanuela Merelli$^1$%
     \email{Emanuela Merelli - emanuela.merelli@unicam.it}
	,
	Cinzia Nitti$^2$%
     \email{Cinzia Nitti - c.nitti@ospedaliriuniti.marche.it}%
	,
	Aldo Salvi$^2$%
     \email{Aldo Salvi - a.salvi@ospedaliriuniti.marche.it}
}


\address{%
  \iid(1)University of Camerino, School of Science and Technology, Computer Science Division, Camerino, IT\\
  \iid(2)Internal and Subintensive Medicine of Ospedali Riuniti - Ancona, IT \\
   \iid(3)Ayasdi, Inc. Palo Alto, CA 
}%

\maketitle


\begin{abstract}
\subsection*{Background}
Pulmonary Embolism (PE)  is a common and potentially lethal condition.  Most patients  die within the first few hours from the event.  Despite diagnostic advances, delays and underdiagnosis in PE are common. Moreover, many investigations pursued in the suspect of PE result negative and no more than 10$\%$ of the pulmonary angio-CT scan performed to confirm PE confirm the suspected diagnosis. To increase the diagnostic performance in PE, current diagnostic work-up of patients  with suspected acute pulmonary embolism   usually starts  with the  assessment of clinical pretest  probability using plasma d-Dimer measurement  and clinical prediction rules. One of the most validated and widely used clinical decision rules are the Wells and Geneva Revised scores. However, both indices have limitations. We aimed to develop a new clinical prediction rule (CPR) for PE based on a new approach for features selection based on topological concepts and artificial neural network.

\subsection*{Results}
Filter or wrapper methods for features reduction cannot be applied to our dataset: the application of these algorithms can only be performed on datasets without missing data. Alternatively, eliminating rows with null values in the dataset would reduce the sample size significantly and result in a covariance matrix that is singular.  Instead, we applied Topological data analysis (TDA) to overcome the hurdle of processing datasets with null values missing data.  A topological network was developed using the Iris software (Ayasdi, Inc., Palo Alto).  The PE patient topology identified two flares in the pathological group and hence two distinct clusters of PE patient populations.  Additionally, the topological netowrk detected several sub-groups among healthy patients that likely are affected with non-PE diseases.   to be diagnosed properly even though they are not affected by PE, in a next study we will introduce also the survival curves for the patients. TDA was further utilized to identify key features which are best associated as diagnostic factors for PE and used this information to define the input space for a back-propagation artificial neural network (BP-ANN). It is shown that the area under curve (AUC) of BP-ANN is greater than the AUCs of the scores (Wells and revised Geneva) used among physicians.

\subsection*{Conclusions}
The results demonstrate topological data analysis and the BP-ANN, when used in combination, can produce better predictive models than Wells or revised Geneva scores system for the analyzed cohort. The new CPR can help physicians to predict the probability of PE. A web-version of the BP-ANN has been published and it can be tested at the address \textit{http:\textbackslash\textbackslash cuda.unicam.it}

\end{abstract}

\ifthenelse{\boolean{publ}}{\begin{multicols}{2}}{}


\section*{Keywords}
Clinical Prediction Rule (CPR), Pulmonary Embolism, Topological Data Analysis, Artificial Neural Network (ANN), Computer Aided Detection CAD

\section*{Background}
\subsection*{A novel computer-aided diagnosis for Pulmonary Embolism}
Several statistical and machine learning techniques have been proposed in the literature to deal with output of implicit or explicit rules and good classification performance \cite{1471-2105-14-S7-S12,1331507}. Most available techniques, such as linear discriminant approaches, multilayer perceptrons or support vector machines, are able to achieve a good degree of provisional accuracy but these methods lack accuraccy sufficient for the implementation of computer-aided diagnosis (CAD) for pulmonary embolism diagnosis. Different studies have been developed for CAD system predicting development of PE in patients. Tang et al used data from the Shangai Xin Hua Hospital, Tourassi et al and Patil S used the data collected from the collaborative study of the \textit{Prospective Investigation of Pulmonary Embolism Diagnosis (PIOPED)} and built neural network \cite{Tourassi1, Patil1}.  
To improve the performance of a CAD we built a new system based on topological data analysis and statistical approach for features selection to define the input space for theartificial neural network. The result of the CAD must be interpreted as a new CPR that might be used to assign a probability of an occurrence of PE\cite{5639424}.

\subsection*{Pulmonary Embolism}
Pulmonary embolism (PE) is a relatively common and potentially lethal condition, affecting a proportion between 0.04$\%$ and 0.09$\%$ of the general population\cite{Tagalakis,Silverstein}, and ranging, in most of the cohorts of outpatients with suspected PE, between 8$\%$ and 12$\%$\cite{Wells,Wolf}. Among patients who die of PE, the largest part is observed in the first few hours from the acute event \cite{Wood}. Despite diagnostic advances, delays in PE diagnosis are common and represent an important issue \cite{Kline}. As a cause of sudden death, PE is second only to arrhythmic death. Among survivors, recurrent embolism and death can be prevented with prompt diagnosis and therapy. Current clinical guidelines suggest to perform a first-level, clinical stratification based upon patient’s history, clinical findings and, in some cases, physician’s judgment. For this purpose, several clinical prediction rulers (CPR) have been suggested and validated, and are currently in use in common clinical practice\cite{Wells,Wolf,Torbicki}. Subsequent management of the patient relies mainly on this stratification. Patients at high-risk for PE should immediately undergo a computed tomography pulmonary angiography (CTPA), while patients with intermediate or low pretest probability should be tested for high-sensitive dDimer assay\cite{Torbicki}. Only patients with increased dDimer levels should be further investigated with second level examinations. CTPA is currently the most widely accepted imaging method recommended to confirm a suspect diagnosis of PE. However, increasing evidences and its direct and indirect costs suggest a limitation in its use: CTPA has been associated to increased risk of secondary cancer from radiation exposure\cite{Einstein} and contrast-induced nephropathy\cite{Mitchell}. Another interesting point is that current use of 64-slices detectors has increased the frequency of subsegmental PE diagnoses: this finding, in the absence of a documented deep vein thrombosis (DVT), may cause clinical uncertainty, and lead to unnecessary therapies\cite{Brunot}. Echocardiography is currently recommended for shocked patients with high suspect of PE. For haemodinamically stable patients it is currently recommended only to better stratify the prognosis by detecting right ventricle dysfunction (RVD)\cite{Torbicki}. Biomarkers such as BNP and troponin are used to assess disease severity and associated to right ventricle overload or damage, but are currently not included in diagnostic algorithms. 

\subsubsection*{Clinical Prediction Rules}

The extreme aspecificity and the variety of the clinical presentation and symptoms and the slenderness of the clinical signs of PE are cause on a side of the proven underdiagnosis on the other of an excess of negative examinations performed. The positive CTPA rate is very low - less than 15$\%$ of the overall number of exams performed for suspected acute PE \cite{Haap}. Recent studies have demonstrated the safety of rejecting the diagnosis of PE by the combination of a low clinical probability, assessed by a CPR, and a normal quantitative d-Dimer test result, thereby decreasing the need for further diagnostic radiological imaging in up to 30$\%$ of patients \cite{Naturwissenschaften2011}. The most widely used CPRs are the Wells rule (table \vref{tab:wells}) and the Geneva Revised score \cite{Wood} The Wells score has three different risk categories, and more recently it has been simplified assuming two risk categories with a cutoff set to 4. Perrier et al has been successful in generating a model based solely on objective parameters using the Geneva Revised score \cite{Torbicki}. This CPR is easily standardizable and has been validated internally and externally, although less extensively than the Wells rule. Both scores appeared to have a comparable predictive value for PE. Regardless  of the rule being performed,  the  proportion of patients with  PE  is around  10$\%$ for low probability, 30$\%$ for moderate probability and  65$\%$ for high clinical probability category.  It is important to note that these scores have severe limitations. The Wells rule includes the physician’s judgment of whether an alternative diagnosis is more likely than PE \cite{Kline,Torbicki} This criterion, which carries a major weight in the score, is subjective and cannot be standardized. Moreover, it has been suggested that the predictive value of the Wells rule is derived primarily from its subjective component \cite{Sanchez} . The Geneve Revised score is based on 13 entirely objective variables, requires a blood gas analysis while breathing room air.  Interestingly, these parameters have only been evaluated for patients in ED with a clinical performance result not superior to the Wells score.

\subsubsection*{Topological data analysis}

Topological Data Analysis (TDA) is a method to analyze multidimensional, complex data primarily driven by geometry. TDA is a result of over a decade of research in applications of pure mathematics to practical problems. The main idea of this approach is that the shape of the data in an abstract multidimensional space drives the analysis by exploring the parallelism of a large number of machine learning algorithms. The three fundamental concepts of TDA are independence of coordinate systems; insensitivity to deformation; and compressed representation \cite{ayasdi}. A typical example of insensitivity to deformation would be writing in a different font as long as the underlying meaning is preserved, while compressed representation refers to approximating a complex shape such that of a circle with a hexagon. Using a mathematical concept of lenses \cite{SPBG07:91-100:2007}, data can be projected onto a subspace suitable for visualization. The topological features of the subspace are then inspected with traditional statistical approaches such as Kolmogorov-Smirnov or t-test analysis. We applied TDA to the clinical data of patients suspected of high risk of PE and used shapes of the generated topological networks on page 16 to identify different subgroups of patients and features that statistically significantly explain the observed differences. We used the TDA-derived features as input into artificial neural network analysis.

\subsubsection*{Artificial Neural Network} 

A complex system may be decomposed into simpler elements, in order to be able to understand it. Also simple elements may be gathered to produce a complex system in line with the approach \textit{divide et impera}. Networks are one approach for achieving this. There are a large number of different types of networks, but they all are characterized by the following components:
\begin{itemize}
\item a set of nodes
\item connections between nodes.
\end{itemize}
The nodes can be treated as computational units. They receive inputs, and process them to obtain an output. The connections determine the information flow between nodes. The interactions of nodes though the connections lead to a global behavior of the network, which cannot be observed in the elements of the network. This global behavior is said to be emergent, meaning that the abilities of the network supersede the ones of its elements. Networks can be used as very powerful tool because many systems can be assembled into a systems-like network space, applications including proteins, computers, and communities. One type of network uses nodes as artificial neurons, and an artificial neuron is a computational model inspired by the natural neurons. Natural neurons receive signals through synapses located on the dendrites or membrane of the neuron. When the signals received are strong enough (greater than a certain threshold), the neuron is activated and emits a signal though the axon. This can activate a cascade process.  The complexity of real neurons is highly abstracted when modeling artificial neurons. These basically consist of inputs, which are multiplied by weights, and then computed by a mathematical function that determines the activation of the neuron. Another function computes the output of the artificial neuron (sometimes in dependence of a certain threshold). Weights can also be negative, so we can say that the signal is inhibited by the negative weight. Depending on the weights, the computation of the neuron will be different. By adjusting the weights of an artificial neuron we can obtain the output we want for specific inputs. If we scale an ANN to hundreds or thousands of neurons, it is both complicated and labor intensive to manually discover all the necessary weights. However, by identifying algorithms to adjust the weights of the ANN in order to obtain the desired output from the network. This process of adjusting weights is called learning or training \cite{Tourassi1} \cite{1165576} \cite{Rosenblatt}.The number of ANN types and uses is very high; hence, there are hundreds of different models considered as ANNs. The differences among them can be related to functions, the accepted values, the topology of network, and/or the learning algorithms, etc. 
We applied backpropagation algorithm to perform a layered feed-forward ANNs. This allows us to organize the artificial neurons in layers to send their signals “forward”, and propagate errors backwards. The network receives inputs by neurons in the input layer, and the output of the network is given by the neurons on an output layer. There may be one or more intermediate hidden layers. The backpropagation algorithm uses supervised learning, which means that we provide the algorithm with examples of the inputs and outputs we want the network to compute, and then the error (difference between actual and expected results) is calculated. The benefit of using backpropagation algorithm is that this error is reduced until the ANN learns the training data. The training begins with random weights, and the goal is to adjust them so that the error will be minimal. Further mathematical explanations are detail in\cite{Rojas1996} and \cite{Kauffman1993}.

\subsection*{Statistical concepts}
\subsubsection*{Kolmogorov-Smirnov Test}
In statistics, the Kolmogorov–Smirnov test (K–S test) is a nonparametric test for the equality of continuous, one-dimensional probability distributions that can be used to compare a sample with a reference probability distribution (one-sample K–S test), or to compare two samples (two-sample K–S test). The Kolmogorov–Smirnov statistic quantifies a distance between the empirical distribution function of the sample and the cumulative distribution function of the reference distribution, or between the empirical distribution functions of two samples. The null distribution of this statistic is calculated under the null hypothesis that the samples are drawn from the same distribution (in the two-sample case) or that the sample is drawn from the reference distribution (in the one-sample case). In each case, the distributions considered under the null hypothesis are continuous distributions but are otherwise unrestricted. The two-sample KS test is one of the most useful and general nonparametric methods for comparing two samples, as it is sensitive to differences in both location and shape of the empirical cumulative distribution functions of the two samples \cite{smirnov}.

\subsubsection*{Receiver Operating Characteristic}
Both Artificial Neural Networks and the two scores (Wells and Revised Geneva) were evaluated by using receiver-operating characteristic (ROC) analysis. Medical tests play a vital role in modern medicine not only for confirming the presence of a disease but also to rule out the disease in individual patient. A test with two outcome categories such as \textit{test +} and \textit{test -} is known as dichotomous, whereas more than two categories such as positive, indeterminate and negative called polytomous test. The validity of a dichotomous test compared with the gold standard is determined by \textit{sensitivity} and \textit{specificity}\cite{Hanley}.
The \textit{Receiver Operating Characteristic (ROC)} curve is the plot that displays the full picture of trade-off between the sensitivity (true positive rate) and (1-specificity) (false positive rate) across a series of inherent validity of a diagnostic test. This curve is useful in:
\begin{itemize}
\item evaluating the discriminatory ability of a test to correctly pick up diseased and non-diseased subjects;
\item finding optimal cut-off point to least misclassify diseased and non-diseased subjects;
\item comparing efficiency of two or more medical tests for assessing the same disease;
\item comparing two or more observers measuring the same test
\end{itemize}
Total area under ROC curve is a single index for measuring the performance a test. The larger AUC - Area Under Curve - , the better is overall performance of the medical test to correctly identify diseased and non-diseased subjects. Equal AUCs of two test represent similar overall performance of tests but this does not necessariy mean that both the curves are identical \cite{McClish1989}.
\subsubsection*{Jaccard Similarity Coefficient}
The Jaccard Similarity coefficient is a statistic used for comparing the similarity and diversity of sample sets. The Jaccard coefficient measures similarity between sample sets, and it is defined as\cite{jaccard}: 
\begin{equation}
J=\frac{|DD \cap AD|}{|DD \cup AD|}
\end{equation}
where:
\begin{itemize}
\item \textit{Doctor diagnosis} -$DD$.
\item \textit{ANN diagnosis} - $AD$.
\end{itemize}

\section*{Results}
\subsection*{Topological data analysis results}
Topologial data analysis was applied on the PE dataset to generate the network in figure (see \vref{fig:IrisNet}) using AYASDI-Iris (Ayasdi, Inc, Palo Alto). We color in red the pathological patients and in blue the healthy group. IRIS highlighted two flares and a longer tail in the pathological group. From the comparison of the two flares we selected a sub-set of features characterized with high Kolmogorov- Smirnov and low p-value (see table \vref{tab:extracted}). The information extracted from the comparison of the two flares is that there is a sub-group of patients who is characterized by the recurrence risk to be affected by PE. IRIS detected some sub-groups among healthy patients, which present other diseases to be diagnosed properly even though they are not affected by PE.

\subsection*{Comparison of AUC: Wells, Geneva and ANN}
From the patients' dataset we random selected 152 patients: 101 pathological and 51 healthy and we studied the discriminatory ability of the system. 
Tables 5 and 6 \vref{tab:t1} show the performance of three classifier: Wells score, revised Geneva score and the ANN model. Any of the three AUC was statistically significant different. The comparative study of three AUCs has been made with the software MedCalc for Windows, version 9.5.0.0 (MedCalc Software, Mariakerke, Belgium)\cite{medcalc}. The AUCs of Revised Geneva and Wells score obtained from the analysed dataset are directly comparable with the results previously published in literature by other groups\cite{Penaloza}.
Figure (see \vref{fig:ann_roc} and \vref{fig:wells_roc} ) showed the ROC curves of  three classifiers, they were evaluated by leave-one-out method\cite{Picard}. The difference between the AUC of the new classifier both with revised Geneva's AUC and Wells' AUC was statistically significant at the $95\%$ confidence level, $P-value < 0.0001$ and $P-value = 0.0025$ respectively. Instead, the difference between revised Geneva's AUC and Wells' AUC is not statistically relevant: $P-value = 0.1456$. Thus, the network's discriminant power is significantly. The Jaccard coefficient for the ANN is $J=0.88$ as every informatics system also the ANN-classifier is affected by the round problem, i.e. the probability value of occurrence of PE in range [0.45; 0.5) is rounded to 0.5 then the patient is classified as pathological. At the state of the art we left to the physician the final judgment over this situations \cite{delong, griner, metz, zweig}.

\section*{Conclusions}
The study derives a new CPR for pulmonary embolism and evaluating patients' cohort characteristic with a topological approach. The new CPR has been obtained training an artificial neural network on the input formed by a set of features selected by Iris. Iris extracted new knowledge from the patients' dataset by the application of an innovative approach for data analysis : \textit{Topological Data Analysis}. Our results show that the feature selection strategy is beneficial for the performance improvement of an ANN trained on the analyzed cohort. A three-layer neural network can be trained to successfully perform the diagnostic task. In conclusion a system based on Iris and an ANN can form the basis of a CAD system to assist physicians with the right stratification of patients. A web-version of the BP-ANN has been published and it can be tested at the address \textit{http:\textbackslash\textbackslash cuda.unicam.it}. In the future we will perform a validation of the system both increasing the number of patients in the dataset and using different cohorts, we will perform a comparison study among artificial neural networks and other classification systems. Our dataset allowed us to increase the clinical performance of about 20$\%$ 

\section*{Methods}
\subsection*{Patients and patients' dataset preparation} 
From a cohort of 1500 patients accepted in the department of Intensive and Subintesive medicine from 2009 to 2012 at \textit{Ospedali Riuniti di Ancona}, a total of 987 patients of average age 70 years were included. The inclusion criteria for this study were that the patients at least were recorded with Wells score, Revised Geneva score, d-Dimer and blood gas $PO_{2}$, the cutoff of d-Dimer was 230ng/ml. For each patients 26 variables were collected (see table \vref{tab:dataset}). In these patients, the diagnosis of PE and the exclusion of PE were made by angio-CT. Among these, 793 had PE and 147 did not have PE. Characteristics of the history, objective data from the physical examination and the outcome of d-Dimer analysis were tabulated for each patient.
To improve the interpretability of our data the logarithm transformation has been applied to d-Dimer and WBC (white blood count). In the present work we analyzed only outpatients.
The patients' dataset was processed using topological data analysis with Ayasdi IRIS specifications. Ayasdi IRIS-ready data files are composed of rows and columns. The calculations are performed on a row-by-row basis and the results elucidate the columns (features) that best define and explain particular groups of rows. The Ayasdi Iris data format is a matrix with unique column headers and unique row identifiers. Column headers must uniquely identify the columns. In contrast, the column that uniquely identifies the rows may be in any column position, it does not have to be the first or the last column.

\subsection*{Classical approaches for feature selection} 
Filter or wrapper methods for features reduction can not be applied to our dataset: the application of these algorithms can be done on a dataset without missing data. The elimination of rows with null values from our dataset would reduce extremily the number of samples and the covariance matrix is singular (ill conditioned)\cite{Trefethen}.

\subsection*{Topological Data Analysis: features selection} 
From the patients' dataset we selected only \textit{d-Dimer, Revised Geneva score and Wells score} for the analysis and Variance Normalized Euclidean as metric function (equation \vref{eq:metric}). The Variance Normalized Euclidean metric is utilized when data are comprised of disparate quantities that are not directly comparable. The filters were L-Infinity centrality, which assigns to each point the distance to the point most distant from it, and \textit{final diagnosis}. L-infinity centrality, is defined for a data point x to be the maximum distance from x to any other data point in the set. Large values of this function correspond to points that are far from the center of the data set (equation \vref{eq:distance}). It is also useful to use some filters that are not geometric, in that they do depend on a coordinate representation, such as coordinates in principal component or projection pursuit analysis. The parameters of the configuration were: resolution 60, gain 3.0x, equalized on (see figure \vref{fig:IrisNet}) \cite{ayasdi}.
\begin{equation}
D(h_1,h_2)=\sqrt{\sum_{i=1}^{d}\frac{(h_1(i)-h_2(i))^2}{\sigma_{i}^{2}}}
 \label{eq:metric}
\end{equation}
\begin{equation}
f(x)=\underset{y \in X}{\max} d(x,y)
\label{eq:distance}
\end{equation}

\subsection*{Artificial Neural Network Architecture} 
The neural network used in our study had a three-layer, feed-forward architecture and was trained by using the back-propagation algorithm with the sigmoid activation function. According to this learning scheme, the network tries to adjust its weights so that for every training input it can produce the desidered output. During the training phase, the network is presented with pairs of input-output patterns: supervised learning. It has been shown that this technique minimizes the mean squared error (MSE) between the decider and the actual network output following an inteative gradient search technique. The input of the network is a subset of the patient's dataset formed with all the patients but only with selected features by TDA (see table \vref{tab:extracted}). Specifically our network had a hidden layer with 3 nodes, and an output layer with a single decision node. The network was trained to output 1 if PE was present and 0 if not. In this study, the network's output was interpreted as the probability of PE being present. In addition, input data were scaled to -1 and +1. The learning rate was selected to be 0.5 and the momentum coefficient to be 0.9, the optimal number of iterations (epochs) have been found equal to 660 after the execution of 100 trials with the permutations of rows in the patients' dataset.

\subsection*{Softwares} 
\subsubsection*{IRIS} 
Topological data analysis has been performed using the Iris software \cite{webayasdi}, properties of Ayasdi Inc. Iris uses the concepts of metric and lenses. Metric is a distance or similarity measure between point. Lenses are mathematical functions through which you see the data points. Different lenses emphasizing different aspects of the dataset and different networks will be generated. Lens executes the division of data points in overlapping bins. Through the bins, the data are clustered. Each cluster is represented by a node, default node size is proportional to the number of data points in the cluster. IRIS connects nodes when corresponding clusters have common data points. During IRIS exploration there are two parameters attached to the lens that allow you to control how the data is divided in the bins. One of these is \textit{resolution} which corresponds to the number of bins is put into: 
\begin{itemize}
\item low resolution: coarse view of data;
\item high resolution: detailed view and possibly fragmented into singletons.
\end{itemize}
Another lens parameter is \textit{gain}. Gain controls how much overlap there is between bins. Increasing the gain increases the number of edges in the output. Oversampling highlights relationship within the data. Finaly the commands \textit{equalize} distributes the data points in the bins so that each bin has an equal number of data points.

\subsubsection*{RULEX 2.0} 
The ANN has been implemented by the Rulex software suite \cite{rulex}. The Rulex software, developed and commercialized by Impara srl, is an integrated suite for extracting knowledge from data through statistical and machine learning techniques. An intuitive graphical interface allows to easily apply standard and advanced algorithms for analyzing any dataset of interest, providing the solution to classification, regression and clustering problems.
To build classifiers we used a number of graphical components provided by Rulex 2.0. We utilized Visualization and editing components to visualize and export the confusion matrix, the training and validation sets, the results of the classifier, to access statistical data (e.g. Covering, Error and Relevance). 

\bigskip

\section*{List of abbreviations}
CPR: clinical predictive rule, CAD: computer aided detection, ANN: artificial neural network, TDA: topological data analysis, PE: pulmonary embolism, DVT: deep venous thrombosis, angio-CT: angiography computed tomography, K-S or KS: Kolmogorov-Smirnov test, ROC: receiver operating characteristic, FP: false positive, TN: true negative, FN: false negative, TP: true positive, DD: doctor diagnosis, AD: artificial neural network dianosis, AUC: area under curve, WBC: white blood count, MSE: mean sqared error, 

\section*{Competing interests}
Damir Herman and Tanya Petrossian are employees of Ayasdi, Inc. and hold stock in the company. The authors declare that they have no competing interests.

\section*{Author's contributions}
Emanuela Merelli, Aldo Salvi, Lorenzo Falsetti and Cinzia Nitti andconceived the project, supervised the study. Lorenzo Falsetti wrote the clinical background. Matteo Rucco designed the experimental flowchart, performed the computer experiments and wrote the manuscript.

\section*{Acknowledgements}
\ifthenelse{\boolean{publ}}{\small}{}
We acknowledge the department of Internal and Subintensive medicine of Ospedali Riuniti di Ancona - Torrette for the patients' dataset, the financial support of the Future and Emerging Technologies (FET) programme within the Seventh Framework Programme for Research of the European Commission, under the FET- Proactive grant agreement TOPDRIM (Topology-driven methods for multievel complex systems), number FP7-ICT-31812, Francesco Vaccarino and Giovanni Petri from ISI foundation for their introduction to algebraic topology and complex networks. Impara s.r.l. and Ayasdi Inc. for their software products. A special thanks is extendend to Damir Herman for his support during the TDA. 
 

\newpage
{\ifthenelse{\boolean{publ}}{\footnotesize}{\small}
 \bibliographystyle{bmc_article} 
 \bibliography{bmc_article} }   


\ifthenelse{\boolean{publ}}{\end{multicols}}{}



\section*{Figures}

\subsection*{Figure 1 - Parallel plots and mean difference}
\includegraphics[width=1\textwidth]{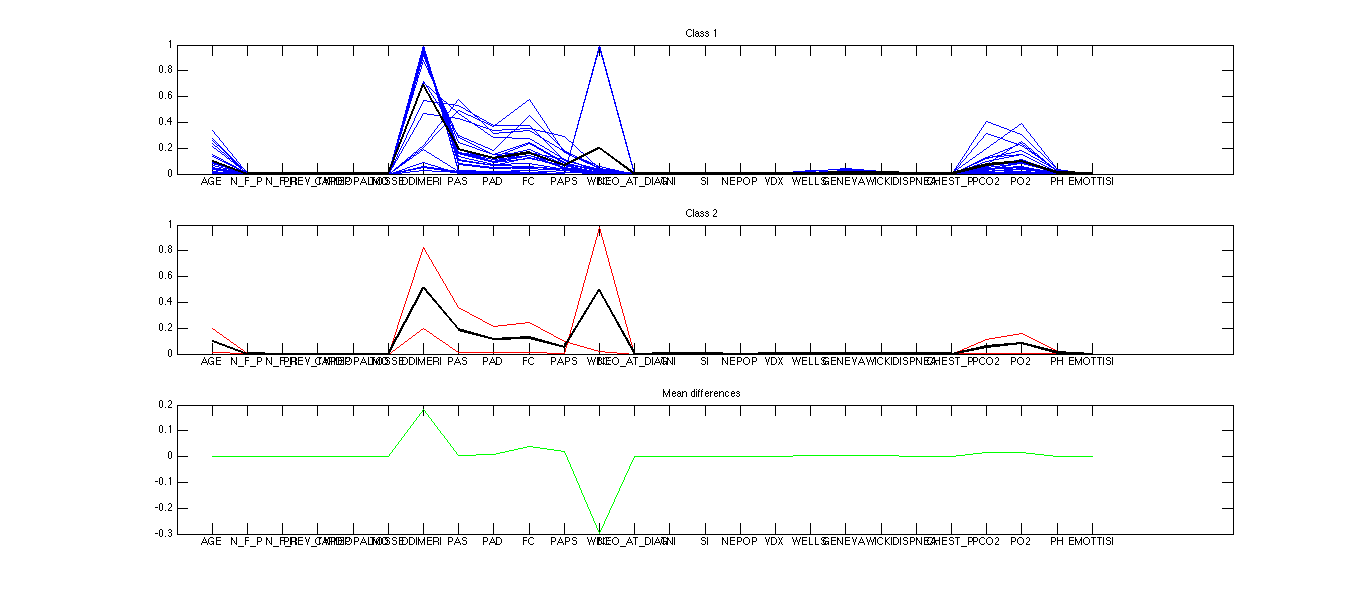}\\
\label{fig:pplot}
Blu lines: parallel plot for pathological patients. Red lines: parallel plot for healthy patients. Green line: difference between mean values of the previous parallel plot; only d-Dimer and WBC exhibit a relevant difference between two classes, but it is not enough to diagnose pulmonary embolism. 

\subsection*{Figure 2 - AYASDI-Iris Analysis}
\includegraphics[width=1\textwidth]{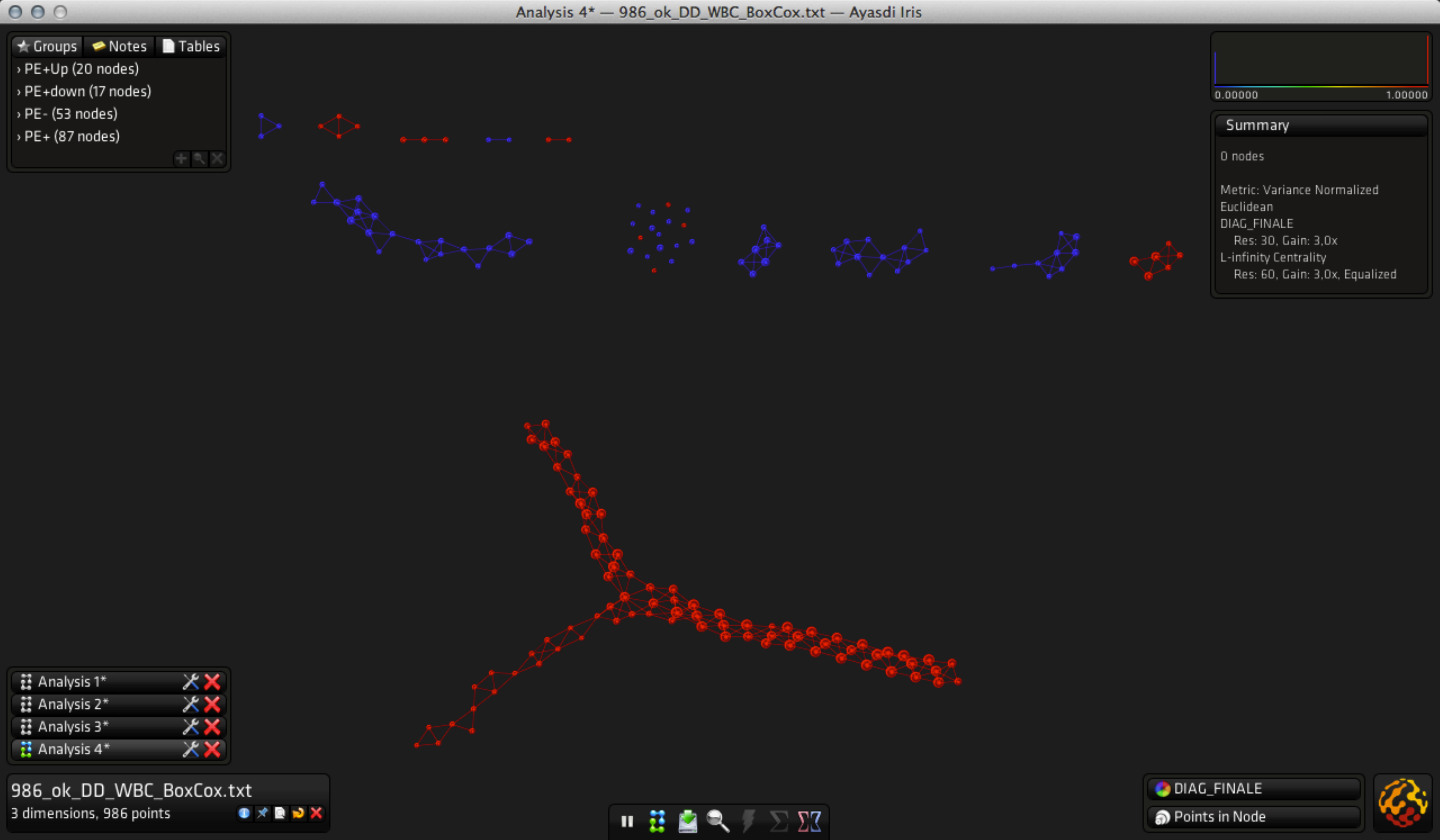}\\
\label{fig:IrisNet}
IRIS highlighted two flares in the pathological group. That means there are two cluster of patients in the PE population (red groups), also IRIS detected some sub-groups among healthy patients, which present other diseases to be diagnosed properly even though they are not affected by PE (blue groups).

\subsection*{Figure 3 - ANN}
\includegraphics[width=0.55\textwidth]{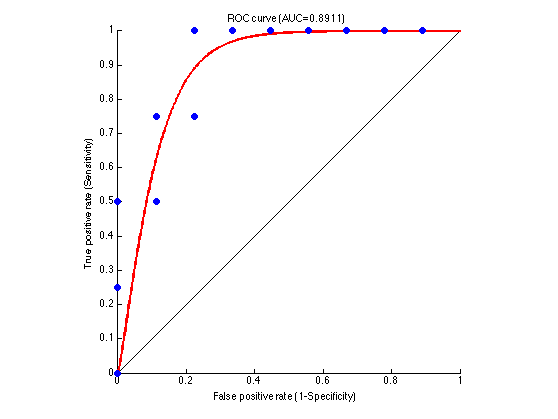}\\ 
\label{fig:ann_roc} 
Comparison study of ROC curves: the AUC obtained with the new classifier is greater then the AUCs from the other two CPRs.

\subsection*{Figure 4 - Revised Geneva}
\includegraphics[width=0.55\textwidth]{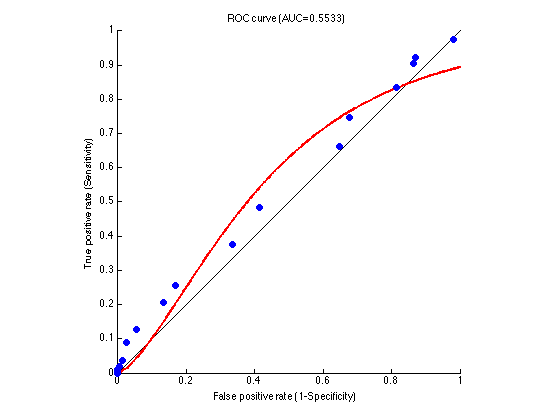}\\ 
\label{fig:geneva_roc} 
ROC curve of Revised Geneva score.

\subsection*{Figure 5 - Score Wells}
\includegraphics[width=0.55\textwidth]{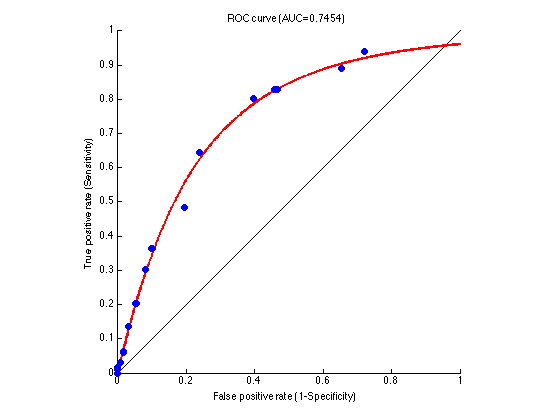}\\ 
\label{fig:wells_roc} 
ROC curve of Wells score.


\section*{Tables}
 \subsection*{Table 1 - Revised Geneva Score}
\par
\mbox{
 \begin{tabular}{|l|l|}
    \hline
    Variable                            & Points     \\ \hline
    \textbf{Predisposing factors}                     & ~       \\ 
    Age $>$ 65 years                         & +1       \\ 
    Previous Deep Vein Thrombosis (DVT) or Pulmonary Embolism (PE) & +3       \\ 
    Surgery or fracture within 1 month               & +2       \\ 
    Active malignacy                        & +2       \\ 
    \textbf{Symptons}                            & ~       \\ 
    Unilateral lower limb pain                   & +3       \\ 
    Haemoptysis                          & +2       \\ 
    \textbf{Clinical signs}                         & ~       \\ 
    Heart rate                           & ~       \\ 
    75-94 beats/min                        & +3       \\ 
    $\geq$ 95 beats/min                      & +5       \\ 
    Pain on lower limb deep vein at palpation an unilateral oedema & +4       \\ 
    \textbf{Clinical probability}                 & \textbf{Total} \\ 
    Low                              & 0-3      \\ 
    Intermediate                          & 4-10      \\ 
    High                              & $\geq$11    \\ 
    \hline
  \end{tabular}
\label{tab:geneva}
}
 \subsection*{Table 2 - Wells Score}
\par
\mbox{
 \begin{tabular}{|l|l|}
    \hline
    Variable                            & Points     \\ \hline
    \textbf{Predisposing factors}                      & ~       \\ 
    Previous Deep Vein Thrombosis (DVT) or Pulmonary Embolism (PE) & +1.5      \\ 
    Recent surgery or immobilization                & +1.5      \\ 
    Cancer                             & +1       \\ 
    \textbf{Symptons}                            & ~       \\ 
    Haemoptysis                          & +1       \\ 
    \textbf{Clinical Signs}                         & ~       \\ 
    Heart rate $>$ 100 beats/min                  & +1.5      \\ 
    Clinical signs of DVT                     & +3       \\ 
    \textbf{Clinical judgement}                       & ~       \\ 
    Alternative diagnosis less likely than PE           & +3       \\ 
    \textbf{Clinical probability (3 levels)}            & \textbf{Total} \\ 
    Low                              & 0 - 1     \\ 
    Intermediate                          & 2 - 6     \\ 
    High                              & $\geq$ 7    \\ 
    \textbf{Clinical probability (2 levels)}            & \textbf{Total} \\ 
    PE unlikely                          & 0 - 4     \\ 
    PE likely                           & $>$4      \\ 
    ~                               & ~       \\
    \hline
  \end{tabular}
\label{tab:wells}
}

\subsection*{Table 3 - Dataset}
 \par
  \mbox{
    \begin{tabular}[t]{|p{1,2cm}|p{3cm}|p{10cm}|}
    \hline
	 & Features & Clinical Significance       \\ \hline
    1 & ID & Patient's identifier       \\ \hline
    2 & Age & With the increase of the age, increase the incidence     \\ \hline
    3 & Number Predictive Factors  & Absolute number of predictive factors   \\ \hline
    4 & Number Risk Factors   & Absolute number of risk factors  \\ \hline
    5 & Previous DVT & A previous DVT / PE is a risk factor repeated infringement DVT / PE \\\hline 
    6 & Palpitations  & Aspecific symptom. If it implies a tachycardia could be associated with DVT / PE  \\ \hline
    7 & Coughs & Symptom very nonspecific but frequently present in patients with DVT / PE      \\ \hline
    8 & d-Dimer & A value of d-Dimer $<230 ng/ml$ is associated with a low risk / absent of DVT / PE. A very high value is associated with a high risk of DVT / PE    \\ \hline
    9 & Systolic Pressure (PAS)  & A low PAS is present in patients with DVT / PE and hemodynamic shock      \\ \hline
    10 & Diastolic Pressure (PAD)  & In cardiogenic shock with DVT / PE is low, sometimes undetectable. By itself has no value despite of the PAS      \\ \hline
    11 & Heart Rate (FC) & In the patient with TVP / EP tachycardia is often found         \\ \hline
    12 & Mean Pulmonary Artery Pressure (PAPS) & It is one of the criteria of right ventricular dysfunction. It can be normal in the case of EP low entity.      \\ \hline
    13 & White Blood Cells Counter (WBC)   & The value increases with inflammatory forms (pneumonia, etc ...) that can be confused with DVT / PE    \\ \hline
    14 & Cancer at diagnosis & It is a risk factor for DVT / EP recognized \\ \hline
    15 & Troponin & It is a marker of myocardial infarction or heart failure and can be confused with DVT / PE    \\ \hline
    16 & Shockindex  & It is the ratio between PAS and FC, if it is greater than 1 is indicative of shock   \\ \hline
    17 & Cancer &  It is a risk factor for DVT / EP recognized   \\ \hline
    18 & Right Ventricular Dsfunction (RVD) & Right ventricular overload in the course of DVT / PE \\ \hline
    19 & Score Wells &    \\ \hline
    20 & Score Revised Geneva  &  \\ \hline
    21 & Score Wicki  &  \\ \hline
    22 & Dyspnea  & Main symptom in DVT / PE    \\ \hline
    23 & Chest pain & Chest pain is present in myocardial infarction, in pleural effusion, in the high DVT / PE    \\ \hline
    24 & pCO2  & Associated with low pO2 may be suggestive of DVT / PE     \\ \hline
    25 & pO2   & Associated with low pCO2 may be suggestive of DVT / PE     \\ \hline
    26 & pH  & In DVT / EP pH is usually normal      \\ \hline
27 & Final Diagnosis  & Final physicians' diagnosis       \\
    \hline
\end{tabular}
 \label{tab:dataset}  	
   }
\subsection*{Table 4 - Extracted Features}
\par
\mbox{
\begin{tabular}[t]{|p{0.2cm}|p{2cm}|p{8cm}|}
    \hline
	 & Features & Clinical Significance       \\ \hline
    1 & Age & With the increase of the age, increase the incidence     \\ \hline
     2 & d-Dimer & A value of d-Dimer $<230 ng/ml$ is associated with a low risk / absent of DVT / PE. A very high value is associated with a high risk of DVT / PE    \\ \hline
    3 & Systolic Pressure (PAS)  & A low PAS is present in patients with DVT / PE and hemodynamic shock      \\ \hline
     4 & Heart Rate (FC) & In the patient with TVP / EP tachycardia is often found         \\ \hline
    5 & Mean Pulmonary Artery Pressure (PAPS) & It is one of the criteria of right ventricular dysfunction. It can be normal in the case of EP low entity.      \\ \hline
    6 & Shockindex  & It is the ratio between PAS and FC, if it is greater than 1 is indicative of shock   \\ \hline
    7 & Score Revised Geneva  &  \\ \hline
    8 & Score Wells  &  \\ \hline
    9 & pO2   & Associated with low pCO2 may be suggestive of DVT / PE     \\ \hline

 \end{tabular}
\label{tab:extracted}
}

 \subsection*{Table 5 - Performance of classifiers}
  
 \par
  \mbox{
   \begin{tabular}{|c|c|c|c|}
    \hline \multicolumn{4}{|c|}{Comparison of classifier}\\ \hline
       & AUC & Standard Error (SE) & $95\%$ Confidence Interval \\ \hline
    ANN & 0,891 & 0,0383 & 0,838 to 0,899 \\ \hline
    Revised Geneva & 0,5533 & 0,0485 & 0,420 to 0,610 \\ \hline
	Wells & 0,7454 & 0,0473 & 0,538 to 0,753\\ \hline

   \end{tabular}
	\label{tab:t1}
   }

\subsection*{Table 6 - Comparison of AUCs}
\par
\mbox{
  \begin{tabular}{|l|l|}
    \hline
    ANN - Geneva        & ~       \\ \hline
    Difference between areas  & 0,337     \\ 
    $95\%$ Confidence Interval & 0,168 to 0,386 \\ 
    Significance level     & $P < 0,0001$ \\
    \hline
    ANN - Wells        & ~       \\ \hline
    Difference between areas  & 0,1456     \\ 
    $95\%$ Confidence Interval & 0,0565 to 0,266 \\ 
    Significance level     & $P = 0,0025$ \\
    \hline
    Revised Geneva - Wells        & ~       \\ \hline
    Difference between areas  & 0,1921     \\ 
    $95\%$ Confidence Interval & 0,0421 to 0,212 \\ 
    Significance level     & $P = 0,1456$ \\
    \hline
  \end{tabular}
  \label{tab:comparison}
}


%

\end{bmcformat}
\end{document}